# Mapping of relativistic Green's functions under extended point canonical transformations

A. D. Alhaidari

*Physics Department, King Fahd University of Petroleum & Minerals, Box 5047, Dhahran 31261, Saudi Arabia*
E-mail: haidari@mailaps.org

Given a relativistic two-point Green's function for a spinor system with spherical symmetry we show how to obtain another in the same class by extended point canonical transformations (XPCT).

## I. THEORY

Let $G(\vec{r},\vec{r}',E)$ be the 4×4 Green's function for a Dirac spinor with relativistic energy $E$. Moreover, let the particle be coupled to the 4-component spherically symmetric potential $[A_0,\vec{A}]=[V(r),\hat{r}W(r)]$ in a non-minimal way as shown in [1,2]. Then, in the atomic units ($m = \hbar = 1$), $G(\vec{r},\vec{r}',E)$ satisfies the following Dirac equation:

$$\begin{pmatrix} 1-E+\alpha^2 V & -i\alpha\vec{\sigma}\cdot\vec{\nabla}+i\alpha\hat{r}\cdot\vec{\sigma}W \\ -i\alpha\vec{\sigma}\cdot\vec{\nabla}-i\alpha\hat{r}\cdot\vec{\sigma}W & -1-E+\alpha^2 V \end{pmatrix} G(\vec{r},\vec{r}',E)=\delta(\vec{r}-\vec{r}')II \qquad (1)$$

where $\alpha$ is the fine structure parameter, $\vec{\sigma}$ are the three 2×2 Pauli spin matrices, $II$ is the 4×4 unit matrix and $\hat{r}=\vec{r}/r$. Spherical symmetry reduces Eq. (1) to the following 2×2 equation for the radial component $g_\kappa(r,r',E)$ of $G(\vec{r},\vec{r}',E)$:

$$\begin{pmatrix} 1-E+\alpha^2 V(r) & \alpha\left(\dfrac{\kappa}{r}+W(r)-\dfrac{d}{dr}\right) \\ \alpha\left(\dfrac{\kappa}{r}+W(r)+\dfrac{d}{dr}\right) & -1-E+\alpha^2 V(r) \end{pmatrix} g_\kappa(r,r',E)=\delta(r-r')I \qquad (2)$$

where $\kappa$ is the spin-orbit coupling parameter defined as $\kappa = \pm (j + ½)$ for $l = j \pm ½$ and $I$ is the 2×2 unit matrix. Eq. (2) can be transformed into its "canonical" form [1-3] using the unitary transformation $U=\exp(i\rho\sigma_2/2)$, where $\rho$ is a real angular parameter, together with the following constraint relating the even and odd components of the relativistic potential:

$$V(r)=\frac{S}{\alpha}\left[W(r)+\frac{\kappa}{r}\right] \qquad (3)$$

The resulting canonical equation for the radial Green's function reads

$$\begin{pmatrix} C-E+2\alpha S\left(W+\dfrac{\kappa}{r}\right) & \alpha\left(-\dfrac{S}{\alpha}+CW+\dfrac{C\kappa}{r}-\dfrac{d}{dr}\right) \\ \alpha\left(-\dfrac{S}{\alpha}+CW+\dfrac{C\kappa}{r}+\dfrac{d}{dr}\right) & -C-E \end{pmatrix} \mathcal{G}_\kappa(r,r',E)=\delta(r-r')I \qquad (4)$$

where $S\equiv\sin(\rho)$, $C\equiv\cos(\rho)$ and $\mathcal{G}_\kappa=Ug_\kappa U^\dagger$.

Now, we are interested in answering the following question [4]:

> Assuming that $\mathcal{G}_\kappa$ is too "difficult" to compute numerically while $\hat{\mathcal{G}}_{\hat{\kappa}}$, which describes another system that belongs to the same class, is not. How can we map $\hat{\mathcal{G}}_{\hat{\kappa}}$ into $\mathcal{G}_\kappa$ using the extended point canonical transformations (XPCT) defined in [3] while treating $\hat{\mathcal{G}}_{\hat{\kappa}}$ as a "black box"?

Since $\mathcal{G}_\kappa$ belongs to the same class as $\hat{\mathcal{G}}_{\hat{\kappa}}$ then, following the formalism developed in [3], we can write $\mathcal{G}_\kappa = R\hat{\mathcal{G}}_{\hat{\kappa}}\bar{R}^{-1}$. Explicitly, this reads,

$$\boxed{\mathcal{G}_\kappa(r,r',E) = R(x)\hat{\mathcal{G}}_{\hat{\kappa}}(x,x',\hat{E})\bar{R}^{-1}(x')} \tag{5}$$

where $r = q(x)$, $R = \begin{pmatrix} g & 0 \\ 0 & h \end{pmatrix}$ and $\bar{R} = \sigma_2 R \sigma_2 = \begin{pmatrix} h & 0 \\ 0 & g \end{pmatrix}$.

$q(x)$, $g(x)$ and $h(x)$ are the transformation functions defined in [3]. Writing Eq. (4) as $(H-E)\mathcal{G}_\kappa(r,r',E) = \delta(r-r')I$ and following the steps in section III of reference [3] we can manipulate this equation as follows:

$$(H-E)\mathcal{G}_\kappa = R\left[R^{-1}(H-E)R\right]\hat{\mathcal{G}}_{\hat{\kappa}}\bar{R}^{-1} = \delta(r-r')I$$
$$= RD^{-1}(\hat{H}-\hat{E})\hat{\mathcal{G}}_{\hat{\kappa}}\bar{R}^{-1} = \delta(r-r')I \tag{6}$$

where the 2×2 nonsingular matrix $D$ is defined in [3] as

$$D = \begin{pmatrix} q'g/h & 0 \\ 0 & q'h/g \end{pmatrix}$$

and $q' = dq/dx$. Thus, we can write Eq. (6) as follows

$$(\hat{H}-\hat{E})\hat{\mathcal{G}}_{\hat{\kappa}}(x,x',\hat{E}) = \delta[q(x)-q(x')]D(x)R^{-1}(x)\bar{R}(x') \tag{7}$$

Now,

$$D(x)R^{-1}(x)\bar{R}(x') = q'\begin{pmatrix} h(x')/h(x) & 0 \\ 0 & g(x')/g(x) \end{pmatrix}$$

and,

$$\delta[q(x)-q(x')] = \frac{1}{dq/dx}\delta(x-x')$$

Hence, Eq. (7) can finally be written as

$$(\hat{H}-\hat{E})\hat{\mathcal{G}}_{\hat{\kappa}}(x,x',\hat{E}) = \delta(x-x')I \tag{8}$$

Therefore, the XPCT given by Eq. (5) maps $\hat{\mathcal{G}}_{\hat{\kappa}}$ into $\mathcal{G}_\kappa$, which is the Green's function for another system in the same class. Stated as an answer to the posed question: "given $\hat{\mathcal{G}}_{\hat{\kappa}}$ which is *not* too difficult to compute then we choose an XPCT (i.e., $q(x)$) that maps it into $\mathcal{G}_\kappa$ of the target problem, which is in the same class but difficult to compute". *It is to be noted that an XPCT that can map two arbitrary Green's functions into each other may not always exist.* A transformation is an XPCT if it keeps the canonical form of the relativistic wave equation invariant. This is equivalent to the statement that $\mathcal{G}_\kappa$ and $\hat{\mathcal{G}}_{\hat{\kappa}}$ belong to the same class.

Aside from the transformation function $q(x)$, we need the parameters of the reference problem (i.e., $\hat{\kappa}, \hat{E}, \hat{S}$ and the parameters of the potential $\hat{W}$) to carry out the calculation of $\mathcal{G}_\kappa(r,r',E)$ in Eq. (5). These parameters are to be evaluated as a function of the target problem parameters ($\kappa$, $E$, $S$ and the parameters of the potential $W$). Eq. (3.3) and Eq. (3.4) in reference [3] are used to achieve that as demonstrated in the Example below. The remaining objects needed to complete the calculation in Eq. (5) are as follows:

$$\begin{aligned} g(x) &= \sqrt{dq/dx} \\ h(x) &= \xi/g(x) \\ \xi &= (\hat{E}+\hat{C})/(E+C) \\ \hat{C} &= \sqrt{1-\hat{S}^2} \end{aligned} \tag{9}$$

## II. EXAMPLE

Let $\hat{\mathcal{G}}_{\hat{\kappa}}$ refer to the Dirac-Oscillator problem [3,5] and $\mathcal{G}_\kappa$ to the Dirac-Coulomb [3,6]. Then, we have the following reference quantities: $\hat{S}=0$, $\hat{C}=1$, $\hat{V}(x)=0$, $\hat{W}(x)=\lambda^2 x$, and $q(x)=x^2$, where $\lambda$ is the oscillator strength parameter. Eq. (3.3) in [3] gives:

$$W(r)=0, \quad V(r)=\frac{S\kappa/\alpha}{r} \equiv \frac{Z}{r}, \text{ and } \hat{\kappa}=2C\kappa+\tfrac{1}{2} \tag{10}$$

resulting in $S=\alpha Z/\kappa$ and $C=\sqrt{1-(\alpha Z/\kappa)^2}$. Putting all of that in equation (3.4) of [3], we finally obtain

$$\begin{aligned} \hat{E} &= \sqrt{1+4\alpha^2\left(\gamma\lambda^2-2ZE\right)} \\ \lambda^2 &= \frac{2}{\alpha}\sqrt{1-E^2} \end{aligned} \tag{11}$$

where $\gamma=C\kappa=\sqrt{\kappa^2-(\alpha Z)^2}$ is the relativistic angular momentum. Therefore, we have all the tools necessary to perform the calculation in Eq. (5) and obtain $\mathcal{G}_\kappa(r,r',E)$.

Specifically, one starts by deciding on the set of numbers $\kappa$, $E$, $Z$, $r$ and $r'$ for which $\mathcal{G}_\kappa$ will be computed. Then Eq. (10) and Eq. (11) together with $x=\sqrt{r}$ will be used to give the corresponding set of numbers ($\hat{\kappa}$, $\hat{E}$, $\lambda$, $x$, $x'$) at which $\hat{\mathcal{G}}_{\hat{\kappa}}$ will be evaluated. Note that an energy dependent expression will substitute for the parameter $\lambda$ in $\hat{\mathcal{G}}_{\hat{\kappa}}$, thus one needs to know $\hat{\mathcal{G}}_{\hat{\kappa}}$ for all $\lambda$. Moreover, the values of $\hat{\kappa}$ obtained by Eq. (10) are not likely to be in the form of integers $\pm 1$, $\pm 2$, ...etc. This has also to be observed in the construction of $\hat{\mathcal{G}}_{\hat{\kappa}}$. Finally, the objects in Eq. (9) will be used to complete the evaluation of $\mathcal{G}_\kappa(r,r',E)$ in Eq. (5).